\documentclass[graphics,floatfix, nofootinbib,tightenlines,nobibnotes, 12pt]{revtex4-2}

\usepackage{graphicx}
\usepackage[english]{babel}
\usepackage[utf8]{inputenc}
\usepackage{amsmath,amssymb}
\usepackage{amsfonts}
\usepackage{multirow}
\usepackage{pstricks}
\usepackage{float}
\usepackage{subfigure}
\usepackage{color}
\usepackage{epsfig}
\usepackage{slashed}
\usepackage[colorlinks=true, linkcolor=blue, citecolor=blue, urlcolor=blue]{hyperref}
\begin{document}
\title{Photoproduction of doubly charmed tetraquark $T_{cc}$ via photon-gluon fusion at ILC and CLIC}

\author{Juan-Juan Niu}
\email{niujj@gxnu.edu.cn}
\author{Hong-Hao Ma}
\email{mahonghao@pku.edu.cn, corresponding author}

\address{Department of Physics, Guangxi Normal University,\\Guilin 541004, People's Republic of China}

\date{\today}

\begin{abstract}
The photoproduction of doubly charmed tetraquark $T_{cc}$ is predicted through the resolved channel $\gamma+g\to \to \langle cc \rangle[n] + \bar{c}+\bar{c} \to T_{cc}+\bar{c}+\bar{c}$ at ILC and CLIC. At $e^{+}e^{-}$ colliders, the initial photons $\gamma$ can be produced from two primary sources, well-delineated within the $Weiz\ddot{a}cker$ Williams approximation (WWA) and the laser back-scattering (LBS). And the initial gluon can be emitted from the photon. The spin and color quantum number $[n]$ of the intermediate diquark configuration can be $\langle cc\rangle[^3S_1]_{\bar{3}}$.
Then its nonperturbative hadronization to $T_{cc}$ was discussed in the phenomenological potential models. Finally, the differential distributions and theoretical uncertainty of the doubly charmed tetraquark $T_{cc}$ were analyzed.
The conclusion is that it is promising to observe $T_{cc}$ via the resolved channel of photoproduction both at the ILC and CLIC, and the results have a strong dependence on the mass of constituent charm quark $m_c$ and the potential model.

\vspace {5mm} 
\end{abstract}

\maketitle

\section{Introduction}
\label{sec:introduction}
%%%%%%%%%%%%%%%%%%%%%%%%%%%%%%%%%%%%%%%%%%%%%%%%%%%%%%%%%%%%%%%%%%%%
The heavy exotic hadrons have attracted increasing attention due to their complex structures, which differ from the structure predicted by the conventional quark models\cite{GellMann:1964nj,Zweig:1981pd,Zweig:1964jf}. The heavy exotic hadrons include tetraquarks, pentaquarks, quark-gluon hybrids, hadron molecular states, etc. Among them, the doubly charmed tetraquark $T_{cc}^{+}(3875)$ was discovered with two heavy charm quarks and two antiquarks ($cc\bar{u}\bar{d}$) and spin-parity quantum numbers $J^{P}=1^{+}$ by the LHCb collaboration in the $D^0D^0\pi^+$ invariant mass spectrum \cite{LHCb:2021auc,LHCb:2021vvq}. In-depth study of the doubly charmed tetraquark is conducive to revealing the internal structure of hadrons and understanding strong interactions. Theoretically, $T_{cc}^{+}(3875)$ has been investigated by multiple methods under molecular state assumptions, such as lattice QCD \cite{Lyu:2023xro,Padmanath:2022cvl,Chen:2022vpo,Collins:2024sfi}, sum rules \cite{Xin:2021wcr}, constituent quark model \cite{Chen:2021tnn,Deng:2021gnb,Deng:2022cld,Ortega:2022efc,Meng:2023jqk,Ma:2023int}, effective field theory \cite{{Fleming:2021wmk,Meng:2021jnw,Chen:2021cfl,Braaten:2022elw,Wang:2022jop,Dai:2023mxm}} and so on \cite{Du:2021zzh,Albaladejo:2021vln,Montesinos:2023qbx,Wang:2023ovj}.
However, as the candidate of compact tetraquark with the $(cc)$-diquark configuration \cite{Zhang:2021yul,Qin:2020zlg,Jin:2021cxj,Weng:2021hje,Guo:2021yws,Agaev:2021vur,Albuquerque:2022weq,Gao:2020bvl,Maiani:2022qze,Liu:2023vrk,Meng:2023for,Noh:2023zoq,Wu:2022gie,Mutuk:2023oyz,Wang:2024vjc,Li:2023wug}, $T_{cc}$ has also been phenomenologically studied through hadronic production \cite{Qin:2020zlg,Chen:2011jtl,Yang:2024ysg}, photon-photon fusion \cite{Jiang:2024lsr}, electron-positron production \cite{Hyodo:2012pm}, and indirect production through Higgs and weak boson decay \cite{Niu:2024ghc,Martynenko:2025nsg}. 

High-energy colliders provide convenient experimental conditions for studying the production and decay properties of the heavy doubly charmed tetraquark states. Among them, the electron-positron Colliders, such as the International Linear Collider (ILC) \cite{ILC:2007oiw,ILC:2007bjz} and the Compact Linear Collider (CLIC) \cite{Linssen:2012hp,CLICdp:2018cto}, have become one of the good platforms for studying exotic states because of their clean background and high luminosity. Relying on the designed collision energy and experimental parameters, we can analyze the contribution of $T_{cc}$ produced through the resolved photon channel, and make a comparative analysis with the results produced by the photon-photon fusion channel to provide certain guidance for the detection of future experiments. At $e^+ e^-$ collider, there are two primary sources of the photon, bremsstrahlung \cite{Frixione:1993yw} or the laser back-scattering (LBS) \cite{Ginzburg:1981vm}. Therefore, based on these two photon sources, we consider $T_{cc}$ as a compact tetraquark and study its production rates via photon-gluon fusion at ILC with center-of-mass (CM) energy $\sqrt{s}=$ 250, 500, 1000 GeV and CLIC with $\sqrt{s}=$ 380, 1500, 3000 GeV, respectively. 
%\cite{Petrelli:1997ge}

Nonrelativistic QCD (NRQCD) \cite{Bodwin:1994jh,Petrelli:1997ge} theory is applied to describe the production of compact tetraquark $T_{cc}$, which can be factorized into short-range coefficients and long-range matrix elements. Specifically, the short-range coefficients are perturbatively obtained by the production of heavy $\langle cc\rangle$-diquark states through the resolved photon channel. The long-range matrix elements describe the  hadronization of the $\langle cc\rangle$-diquark to $T_{cc}$. This process is nonperturbative and highly dependent on the potential model, capable of describing the key information of heavy exotics.
The spin and color quantum number of the intermediate $\langle cc\rangle$-diquark can be $[^3S_1]_{\mathbf{\bar{3}}}$ and $[^1S_0]_6$ for the decomposition of the $\rm SU(3)_C$ color group, $\mathbf{3} \otimes \mathbf{3}  = \mathbf{\bar{3}} \oplus \mathbf{6}$. 
In this paper, only the $[^3S_1]_{\mathbf{\bar{3}}}$ configuration is taken into account to form a compact diquark state. 

The rest parts of this paper are arranged as follows. 
In Sec. \ref{Calculation}, the calculation technology is introduced based on NRQCD. 
Then we present the numerical results of the cross sections, the differential distributions for the production of $T_{cc}$, and the main sources of theoretical uncertainty in Sec. \ref{results}. Finally, Sec.\ref{summary} is reserved for a summary.

%%%%%%%%%%%%%%%%%%%%%%%%%%%%%%%%%%%%%%%%%%%%%%%%%%%%%%%%%%%%%%%%%%%%
\section{Calculation Technology}
\label{Calculation}
%%%%%%%%%%%%%%%%%%%%%%%%%%%%%%%%%%%%%%%%%%%%%%%%%%%%%%%%%%%%%%%%%%%%

Within the framework of NRQCD, the differential cross section for the production of doubly charmed tetraquark $T_{cc}$ via photon-gluon fusion in $e^+ e^-$ collision can be factored into
\begin{eqnarray}
    d\sigma (e^+e^-\to e^+e^-+T_{cc}+\bar{c}+\bar{c})=\int dx_{1}f_{\gamma/e}(x_{1})\int dx_{2} f_{\gamma/e}(x_{2}) \nonumber\\
    \times\sum_{i,j}\int dx_{i}f_{i/\gamma}(x_{i})\int dx_{j} f_{j/\gamma}(x_{j})
    \times \sum_n d \hat{\sigma}( ij\to \langle cc\rangle [n]+\bar{c}+\bar{c})\langle \mathcal{O}_{[n]}^{T_{cc}}\rangle.
\label{factorization}
\end{eqnarray}
Here $f_{\gamma/e}(x)$ represents the photon distribution function, where $x=E_{\gamma}/E_{e}$ is the energy fraction of the photon emitted from the initial electron or positron, $f_{i/\gamma}(i=\gamma,g,u,d,s)$ is the Gl$\ddot{\rm u}$ck-Reya-Schienbein (GRS) distribution function of parton $i$ in
 photon \cite{Gluck:1999ub} and $f_{\gamma/\gamma}=\delta(1-x)$ is for the direct photoproduction process, $d \hat{\sigma}$ is the differential partonic cross section for hard subprocess $\gamma+g\to \langle cc\rangle[^3S_1]_{\mathbf{\bar{3}}}+\bar{c}+\bar{c}$,
$\langle \mathcal{O}_{[n]}^{T_{cc}}\rangle$ is the hadronization from $\langle cc\rangle$-diquark to $T_{cc}$, and its nonperturbative effect usually depends on the potential model.
 
At $e^+e^-$ colliders, there are two main sources that describe the distribution function of initial photon $f_{\gamma/e}(x)$. One of them is emanated through the photon bremsstrahlung of the initial electrons or positrons, which can be well described by the Weizsacker-Williams approximation (WWA) \cite{Frixione:1993yw},
\begin{equation}
    f^{\mathrm{WWA}}_{\gamma/e}(x) = \frac{\alpha}{2\pi}\left(\frac{1+(1-x)^2}{x}\log\left(\frac{Q^{2}_{\rm max}}{Q^{2}_{\rm min}}\right)+2m^2_{e}x\left(\frac{1}{Q^{2}_{\rm max}}-\frac{1}{Q^{2}_{\rm min}}\right)\right),
    \label{eq:wwa}
\end{equation}
where $\alpha$ denotes the electromagnetic fine structure constant, $Q^{2}_{\rm min}=m^{2}_{e}x^{2}/(1-x)$, $Q^{2}_{\rm max}=(\theta_{c}\sqrt{s}/2)^2(1-x)+Q^{2}_{\rm min}$, $\sqrt{s}$ is the CM energy, and $\theta_{c}=32$ mrad is the maximum scattering angle of the electron or positron to ensure the real photon \cite{Klasen:2001cu}. 
Another source of photon is the laser back-scattering (LBS) and its photon energy spectrum is \cite{Ginzburg:1981vm},
\begin{equation}
    f^{\mathrm{LBS}}_{\gamma/e}(x)=\frac{1}{N}\left(1-x+\frac{1}{1-x}-\frac{4x}{x_{m}(1-x)}+\frac{4x^2}{x_{m}^2(1-x)^2}\right),
    \label{eq:lbs}
\end{equation}
where the normalization factor $N$ is
\begin{equation}
    N=\left(1-\frac{4}{x_{m}}-\frac{8}{x^{2}_{m}}\right)\log(1+x_{m})+\frac{1}{2}+\frac{8}{x_{m}}-\frac{1}{2(1+x_{m})^2},
\end{equation}
$x_m=4E_e E_l \rm cos^2\frac{\theta}{2}$ with the energy of
incident electron and laser beams $E_e$ and $E_l$, $\theta$ is the angle between them, and the energy fraction $x$ of LBS is restricted by 
\begin{equation}
0 \leq x \leq \frac{x_m}{1+x_m}~(x_{m} \approx 4.83 \cite{Telnov:1989sd}).
\end{equation}

We adopt both the WWA and LBS photon distribution functions for the photon at ILC with $\sqrt{s} = 250, 500, 1000$ GeV and CLIC with $\sqrt{s} = 380, 1500, 3000$ GeV, respectively. 

\subsection{Short-distance Coefficient}

\begin{figure}[!thbp]
\centering
    \includegraphics[scale=0.25]{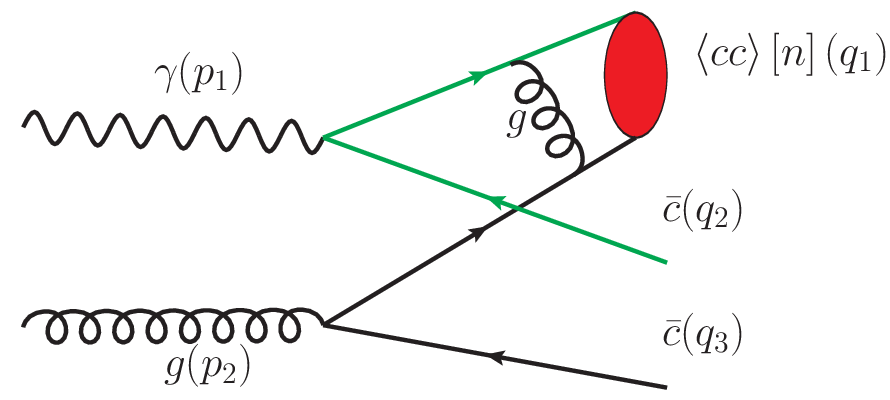}
    \includegraphics[scale=0.25]{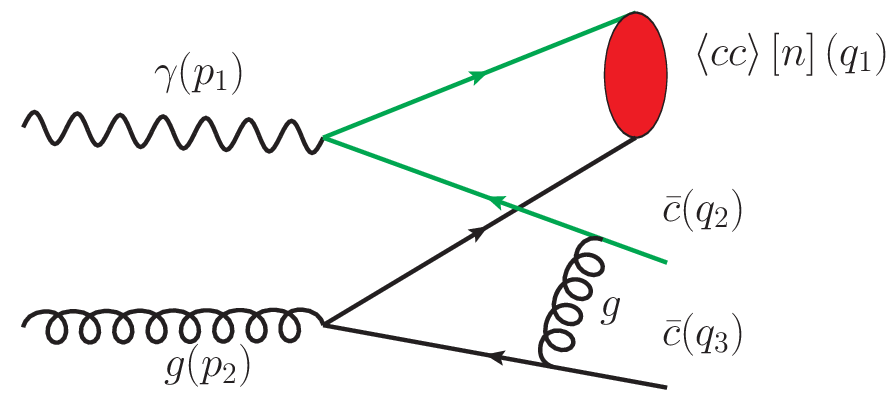}
    \includegraphics[scale=0.25]{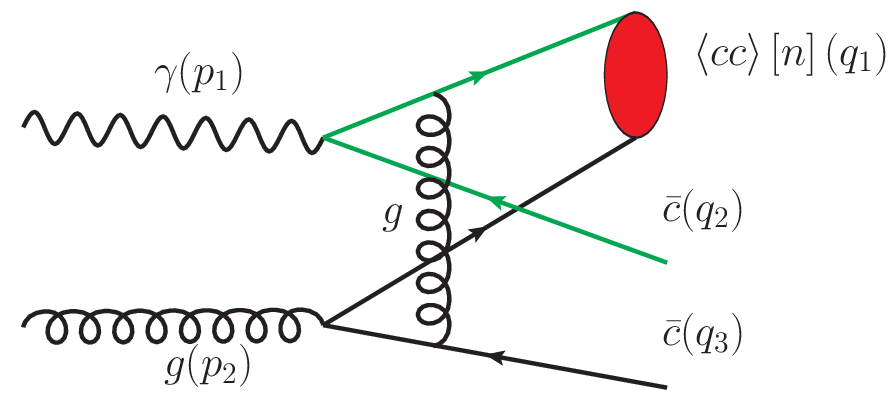}
    \includegraphics[scale=0.25]{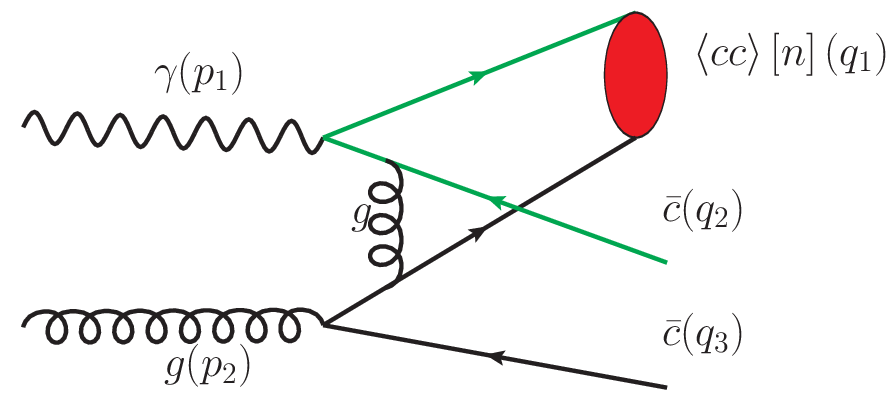}\\
\vspace{0.3cm}    
    \includegraphics[scale=0.25]{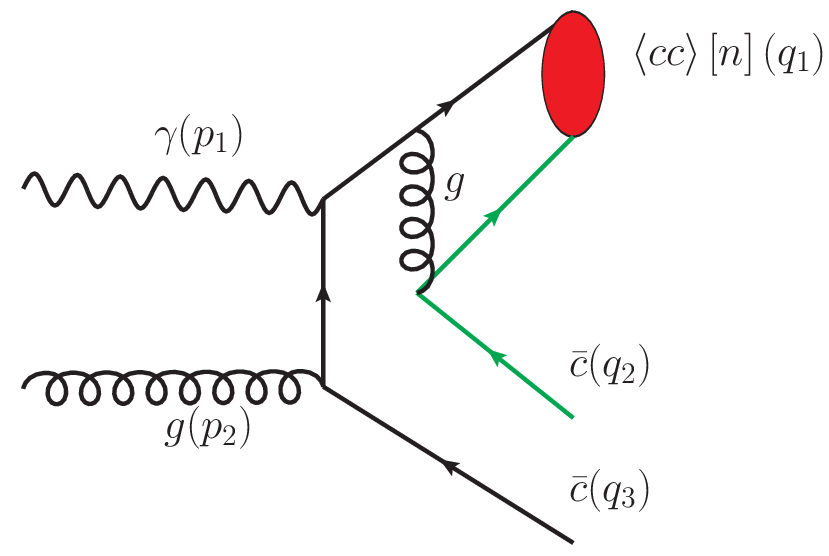}
    \includegraphics[scale=0.25]{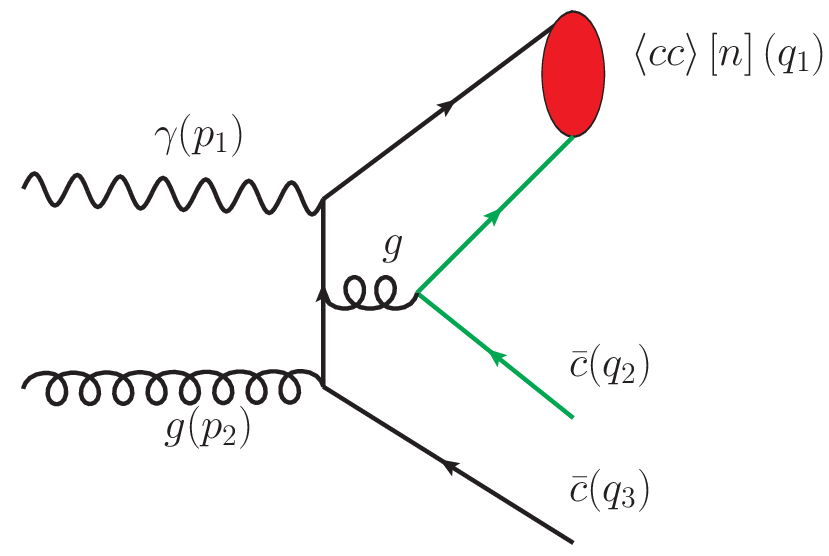}
    \includegraphics[scale=0.25]{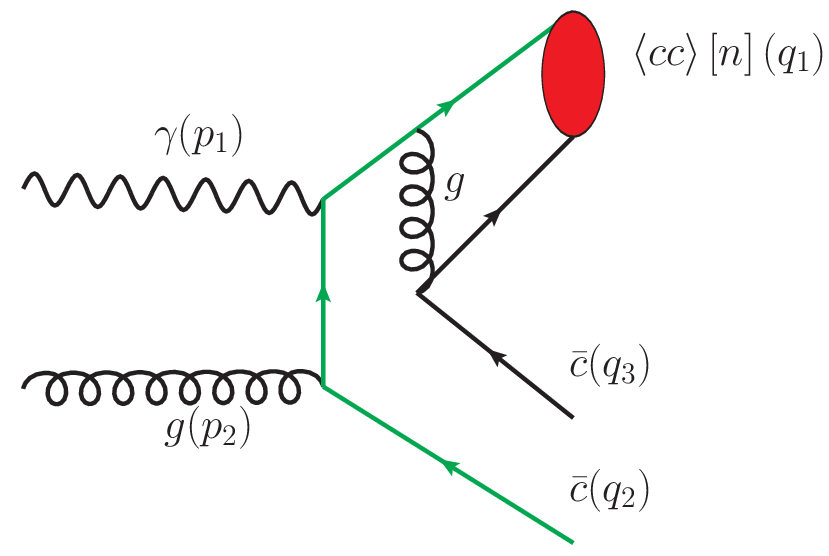}
    \includegraphics[scale=0.25]{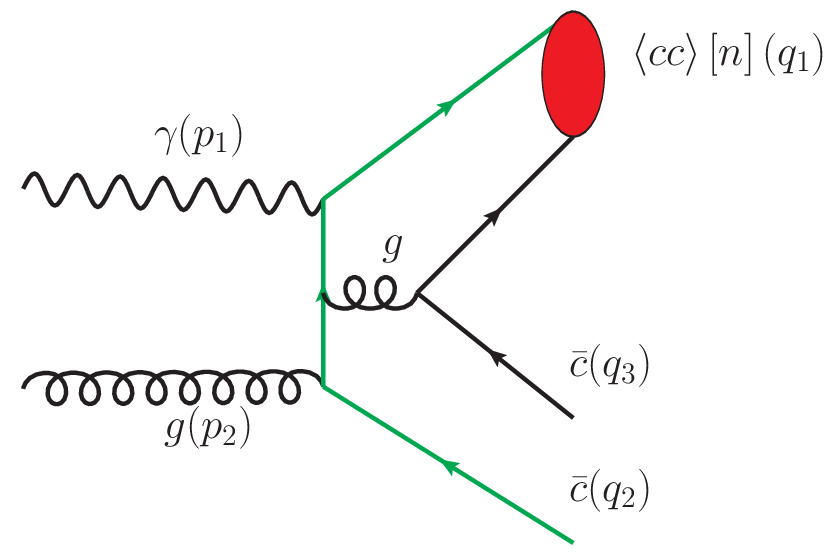}\\
\vspace{0.3cm}        
    \includegraphics[scale=0.22]{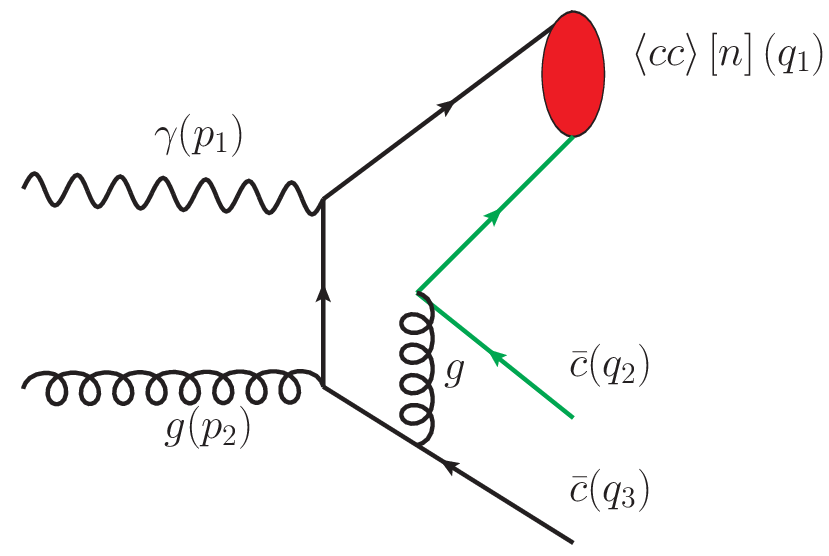}
    \includegraphics[scale=0.22]{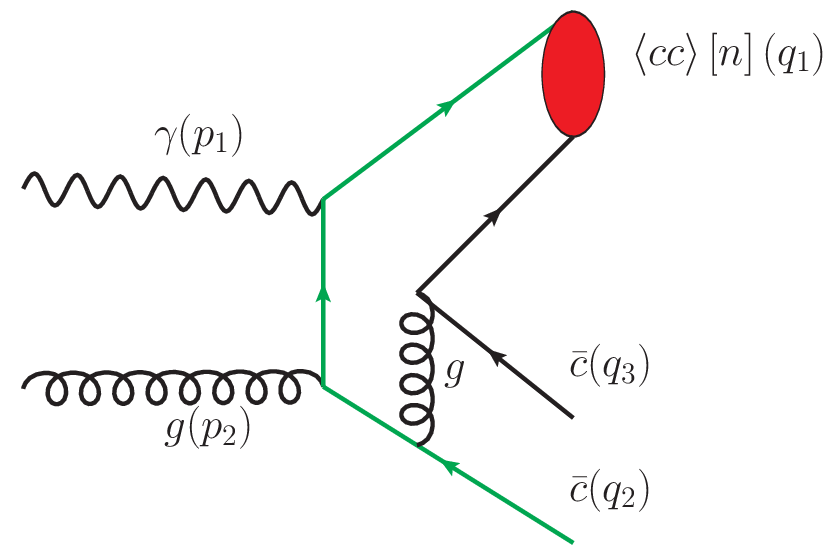}
    \includegraphics[scale=0.25]{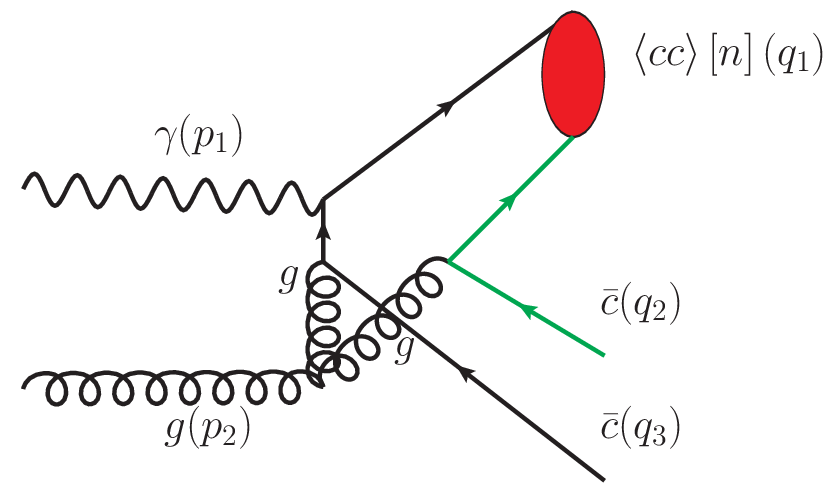}
    \includegraphics[scale=0.25]{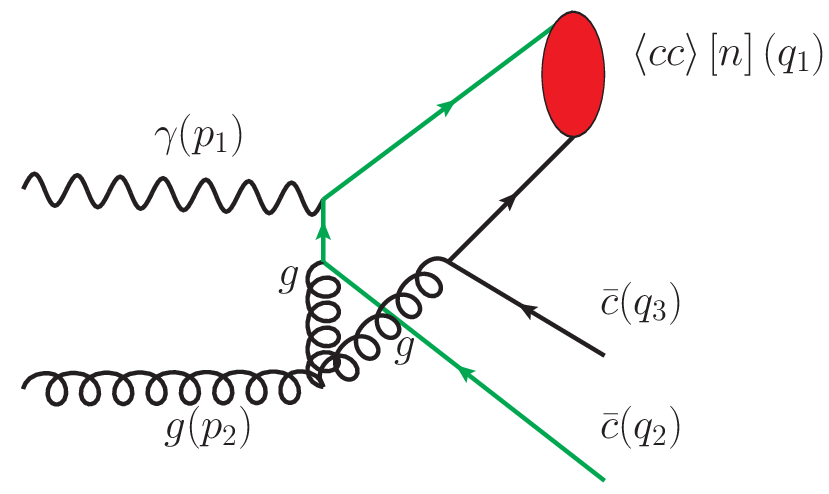}
\caption{Twelve typical Feynman diagrams for $\gamma+g\to \langle cc\rangle[n]+\bar{c}+\bar{c}$ subprocess, and another twelve can be obtained by interchanging the initial photon and gluon lines.} 
\label{feynman}
\end{figure}

Typical Feynman diagrams for the hard partonic process, $\gamma(p_1)+g(p_2)\to \langle cc\rangle [n](q_1)+\bar{c}(q_2)+\bar{c}(q_3)$ at leading order in ${\cal O}(\alpha_s^3)$ are shown in Fig. \ref{feynman}, and another twelve can be obtained by interchanging the initial photon and gluon lines. Here only $[n]=[^3S_1]_\mathbf{\bar{3}}$ configuration is taken into account to obtained the compact $\langle cc\rangle$-diquark state.

The differential cross section $d\hat{\sigma}$ can be rewritten as
\begin{eqnarray}
d\hat\sigma(\gamma+g\to \langle cc\rangle[n]+\bar{c}+\bar{c})=\frac{1}{4\sqrt{(p_1\cdot p_2)^2-m^4_{e}}} \overline{\sum}  \big|{\cal M\rm [n]}\big|^{2} d\Phi_3,
\label{sd-sigma}
\end{eqnarray}
where $\overline{\sum}$ stands for the average over the spins of the initial states and the sum over colors and spins of the final states, $d\Phi_3$ is the three-body phase space.

Note that, an additional factor $2^2/(2!2!)$ shall be multiplied by the squared amplitudes, where $2^2$ contributes to the additional 24 Feynman diagrams resulting from the exchange of identical particles in the $\langle cc\rangle$-diquark state, $1/(2!2!)$ due to the symmetry of identical particles of two heavy quarks inside the diquark and two final-state antiquarks. The momenta of the constituent quark
in $\langle cc\rangle$-diquark can be $q_{11} = q_{1}/2 +q$ and $q_{12} = q_{1}/2-q$.
with the small relative momentum $q$ between two constituent quarks and $q$ = 0 at the leading velocity expanding in NRQCD.

The amplitude $\cal M$ for the production of heavy $\langle cc\rangle$-diquark can be obtained by applying the charge conjugate matrix $C = -i\gamma^{2}\gamma^{0}$ to reverse one of the fermion lines in the Feynman diagrams (green in Fig. \ref{feynman}). After the action of $C$ parity, the amplitude for the production of $\langle cc\rangle$-diquark can be related to the amplitudes of charmonium with an additional factor $(-1)^{\zeta+1}$, where $\zeta$ is the number of vector vertices in the reversed $c\bar{c}$ fermion chain \cite{Jiang:2024lsr,Ma:2025ito}. 

The projector for two constituent charm quarks $v(q_{12})\bar{u}(q_{11})$ into $\langle cc \rangle[^3S_1]_{\bar{3}}$ take the following replacement \cite{Petrelli:1997ge},
\begin{equation}
\label{projector}
\frac{1}{2 \sqrt{2m_c}} \slashed{\epsilon}(q_1)(\slashed{q}_1+2m_c)\otimes {\cal G} ,
\end{equation}
$\epsilon(q_1)$ is the polarization vector of spin-triplet state, and ${\cal G}$ is the color factor of the diquark. Note that the mass of diquark $2m_c$ can be different from the mass $m_{T_{cc}}$ because the subsequent hadronization of diquark can further generat mass.

The color factor $\mathcal C_{ij,k}$ for the production of $\langle cc \rangle[^3S_1]_{\bar{3}}$ has been extracted from the amplitudes and it satisfies the following form
\begin{eqnarray}
{\cal C}_{ij,k}={N}\times\sum_{a=1}^{8}\sum_{m,n=1}^{3}(T^{a})_{im}(T^{a})_{jn} \times {\cal G}_{mnk},
\end{eqnarray}
where ${{N}}=\sqrt{1/2}$ is the normalization constant, $i,j,m,n$ are the color indices of two outgoing antiquarks and two constituent charm quarks of the diquark respectively, $k$ stands for the color index of the diquark, ${\cal G}_{mnk}$ is the color factor for diquark in Eq. \eqref{projector} and it can be adopted the anti-symmetric function $\varepsilon_{mnk}$ for $\langle cc \rangle_{\mathbf {\bar{3}}}$, which satisfies,
\begin{equation}
\varepsilon_{mnk}\varepsilon_{m'n'k} =\delta_{mm'}\delta_{nn'}-\delta_{mn'}\delta_{nm'}.
\end{equation}
Finally, the amplitudes of all 24 Feynman diagrams need to be summed, $\mathcal M\rm [n]=\sum_{i=1}^{24}\mathcal M_{i}\rm[n]$.
%%%%%%%%%%%%%%%%%%%%%%%%%%%%%%%%%%
\subsection{Long-distance matrix element}
\label{LDME}
%%%%%%%%%%%%%%%%%%%%%%%%%%%%%%%%%%
The long-distance matrix element ${\mathcal O}_{\langle cc\rangle[n] }^{ T_{cc}}$ in Eq. \eqref{factorization} describes the hadronization from intermediate $\langle cc\rangle$-diquark in spin and color state $[n]$ to $T_{cc}$, which is accompanied by a large nonperturbative effect and has a high dependence on the potential model. In this paper, we will theoretically predict its value in three different hadronization schemes. First of all, we will simply adopt the nonrelativistic quark model to correlate the long-range matrix element with  Sch$\ddot{\mathrm{o}}$dinger wave function at the origin $\Psi_{T_{cc[n]}}(r=0)$, and then they can be naturally associated with the radiation wave function at origin for S-wave \cite{Hyodo:2012pm},
\begin{equation}
{\mathcal O}_{\langle cc\rangle[n] }^{ T_{cc}} \longrightarrow |\Psi_{\langle cc\rangle[n]}^{T_{cc}}(r=0)|^2 = \frac{|R_{\langle cc\rangle[n]}^{T_{cc}}(r=0)|^2}{4\pi}.
\label{HOP}
\end{equation}
Under the harmonic oscillator potential (HOP) \cite{Hyodo:2012pm}, the solved value of ${\mathcal O}_{\langle cc\rangle[n] }^{ T_{cc}}$ is 0.089 ${\rm GeV}^3$.

After that, the diquark-antiquark symmetry (DAS) \cite{Lichtenberg:1989ix,Anselmino:1992vg,Carlson:1987hh,Savage:1990di,Brambilla:2005yk,Fleming:2005pd,Cohen:2006jg} is utilized to analyze the long-distance matrix element. According to the color interaction, the $\langle cc\rangle$-diquark in color antitriplet can act as an antiquark ${\bar{c}}$, which has the same color. Therefore, the hadronization of $(cc)_{\bar{3}}$ into the tetraquark state $T_{cc}$ with quark constituents $cc\bar{u}\bar{d}$ can be related to the transition of an anti-charm quark $\bar{c}$ to heavy anticharm baryon $\bar{\Lambda}_c^-$ with quark constituents $\bar{c}\bar{u}\bar{d}$, i.e., $(cc)_{\bar{3}}\to T_{cc}(cc\bar{u}\bar{d}) \Longleftrightarrow \bar{c} \to \bar{\Lambda}_c^-(\bar{c}\bar{u}\bar{d})$. And the transition probability of $\bar{c} \to \bar{\Lambda}_c^-(\bar{c}\bar{u}\bar{d})$ can be estimated by the fragmentation fraction of $c \to \Lambda_c^+ (cud)$, whose value can be obtained by fitting the experimental data $f(c \to \Lambda_c^+)=0.06$ \cite{Lisovyi:2015uqa}. For the transition of a heavy $cc$ pair into $\langle cc\rangle$-diquark, the Sch$\ddot{\mathrm{o}}$dinger wave function at the origin $\Psi_{cc}(0)$ is available, its value is slightly different and is estimated in different potential models. Under the Power-law potential \cite{Bagan:1994dy}, $|\Psi_{cc}(0)|^2 = 0.039 \;{\rm GeV}^3$ is estimated in Ref. \cite{Baranov:1995rc}, and directly used in the numerical calculation. Therefore the hadronization ${\mathcal O}_{\langle cc\rangle[n] }^{ T_{cc}}$ can be rewitten as,
\begin{equation}
     {\mathcal O}_{\langle cc\rangle[n] }^{ T_{cc}} \longrightarrow |\Psi_{cc}(0)|^2  \times f(c \to \Lambda_c^+) = 0.00234 \;{\rm GeV}^3.
     \label{HDAS}
\end{equation}

Similarly, DAS symmetry can be used to describe the light antidiquark $\langle \bar{u}\bar{d}\rangle$, the light antiquarks of $T_{cc}$, and its color interaction can be regarded as a light quark \cite{Cheng:2020wxa}. When the heavy $cc$ pair transitions to the $\langle cc\rangle$-diquark by $\Psi_{cc}(0)$, and then a light quark can be captured $100\%$ from the vacuum to form the tetraquark state. This process is the same as the hadronization of doubly charmed baryons. By this way, the hadronization ${\mathcal O}_{\langle cc\rangle[n] }^{ T_{cc}}$ can be
\begin{equation}
     {\mathcal O}_{\langle cc\rangle[n] }^{ T_{cc}} \longrightarrow |\Psi_{cc}(0)|^2 = 0.039 \;{\rm GeV}^3.
\label{DAS}
\end{equation}

These three hadronization schemes for the production of tetraquarks are accompanied by great uncertainties, and we will also conduct in-depth analysis in the numerical part.

%%%%%%%%%%%%%%%%%%%%%%%%%%%%%%%%

\section{Numerical Results}
\label{results}

The input parameters in the numerical calculation are listed below:
\begin{eqnarray}
&&m_c=1.8~\rm{GeV},~~~~~~~~~~|\Psi_{cc}(0)|^2=0.039~{\rm GeV}^3,\nonumber\\
&&\alpha=1/132.23,~~~~~~~~~~~~\mu = \sqrt{4m_{c}^2 + p^2_T},
\end{eqnarray}
$\mu$ is the renormalization scale, which is typically taken as the transverse mass of the final-state doubly charmed tetraquark with its transverse momentum $p_T$. The strong running coupling can be obtained from the solution of the one-loop renormalization group equation with the reference point $\alpha_s(m_Z)$=0.1180 \cite{ParticleDataGroup:2024cfk}. The long-distance matrix element ${\mathcal O}_{\langle cc\rangle[n] }^{ T_{cc}}$ obtained by HDAS hadronization scheme in Sec. \ref{LDME} is used as a representative for numerical calculation, and the theoretical uncertainty caused by the long-distance matrix element obtained in three different hadronization schemes are analyzed subsequently. 

The cross sections $\sigma$ for the production of $T_{cc}[^3S_1]_{\bar{3}}$ at ILC and CLIC are calculated through the resolved photon-gluon fusion using both WWA and LBS photon spectrum, respectively. For comprehension, the cross sections via photon-photon fusion based on the same parameters are all listed in Table \ref{crosssenction}. The designed CM collision energy is selected to be $\sqrt{s}=$ 250, 500, and 1000 GeV at ILC, and $\sqrt{s}=$380, 1500, and 3000 GeV at CLIC.

From Table \ref{crosssenction}, we can see that the contribution from LBS photon spectrum is dominant than WWA. With the increase of CM energy, the trend of the cross section is to rise first and then decrease for LBS photon spectrum. However for WWA photon spectrum, it is not until the CM energy reaches 3000 GeV that its downward trend was shown. For resolved photon-gluon fusion, the cross section keeps showing an increasing trend when the collision energy increases to 3000 GeV. In WWA photon spectrum, as the CM energy increases, the contribution of the resolved photon-gluon channel continues to rise, being 3.53\%, 6.38\%, 12.61\% of that of the photon-photon channel. The same conclusion can also be obtained at CLIC.
While in LBS photon spectrum, the contribution of the resolved photon-gluon channel increased significantly with the increase of CM energy at ILC, up to 88.77\%, 3.53, 14.60 times of that of the photon-photon channel. At CLIC, the contribution of the resolved photon-gluon channel can be 2.04, 32.48, and 19.04 times of that of the photon-photon channel at $\sqrt{s}=$380, 1500, and 3000 GeV.
Therefore, the study of the production of tetraquark states through resolved photon-gluon channel cannot be ignored at $e^+e^-$ collider.

With the designed luminosity $\mathcal L=2, 4, 8~ {\rm ab}^{-1}$ corresponding to ILC at $\sqrt{s}=$ 250, 500, and 1000 GeV and $\mathcal L=1, 2.5, 5~{\rm ab}^{-1}$ for CLIC at $\sqrt{s}=$380, 1500, and 3000 GeV, respectively, the total produced events of $T_{cc}$ via photoproduction can be estimated by $\sigma\times\mathcal L$, approximately $9.37\times 10^3$ ($1.92\times 10^5$), $3.13\times 10^4$ ($4.01\times 10^5$) and $8.43\times 10^4$ ($1.06\times 10^6$) each year at ILC with $\sqrt{s}=$ 250, 500, and 1000 GeV using WWA (LBS) photon spectrum. At CLIC with $\sqrt{s}=$380, 1500, and 3000 GeV, the total produced events of $T_{cc}$ via photoproduction can be $6.58\times 10^3$ ($9.50\times 10^4$), 2.92$\times 10^4$ (3.91$\times 10^5$) and 6.57$\times 10^4$ (1.58$\times 10^5$) each year using WWA (LBS) photon spectrum. Based on such a base number and Considering the subsequent decay of the doubly charmed tetraquark, it is still possible for $T_{cc}$ to be experimentally discovered at linear electron-positron colliders through photoproduction.

\begin{table}[htb]
\caption{The cross sections (in unit: fb) for $T_{cc}[^3S_1]_{\bar{3}}$ production  at ILC and CLIC via photon-photon and photon-gluon fusion using WWA and LBS spectrum.}
\begin{tabular}{|cc|ccc|ccc|}
\hline
\multicolumn{2}{|c|}{\multirow{2}{*}{$\sqrt{s}$ (GeV)}}      & \multicolumn{3}{c|}{ILC}                                      & \multicolumn{3}{c|}{CLIC}                                      \\ \cline{3-8} 
\multicolumn{2}{|c|}{}                                       & \multicolumn{1}{c|}{250}  & \multicolumn{1}{c|}{500}  & 1000  & \multicolumn{1}{c|}{380}  & \multicolumn{1}{c|}{1500}  & 3000  \\ \hline
\multicolumn{1}{|c|}{\multirow{3}{*}{WWA}} & $\gamma+\gamma$ & \multicolumn{1}{c|}{4.53} & \multicolumn{1}{c|}{7.37} & 9.36  & \multicolumn{1}{c|}{6.27} & \multicolumn{1}{c|}{9.81}  & 9.28  \\ \cline{2-8} 
\multicolumn{1}{|c|}{}                     & $\gamma$ + g    & \multicolumn{1}{c|}{0.16} & \multicolumn{1}{c|}{0.47} & 1.18  & \multicolumn{1}{c|}{0.31} & \multicolumn{1}{c|}{1.88}  & 3.86  \\ \cline{2-8} 
\multicolumn{1}{|c|}{}                     & total           & \multicolumn{1}{c|}{4.68} & \multicolumn{1}{c|}{7.84} & 10.53 & \multicolumn{1}{c|}{6.58} & \multicolumn{1}{c|}{11.70} & 13.13 \\ \hline
\multicolumn{1}{|c|}{\multirow{3}{*}{LBS}} & $\gamma+\gamma$ & \multicolumn{1}{c|}{51.04}    & \multicolumn{1}{c|}{22.11}    &   8.52  & \multicolumn{1}{c|}{31.27}    & \multicolumn{1}{c|}{4.67}     &   1.58  \\ \cline{2-8} 
\multicolumn{1}{|c|}{}                     & $\gamma$ + g    & \multicolumn{1}{c|}{45.31}    & \multicolumn{1}{c|}{78.11}    &   124.43   & \multicolumn{1}{c|}{63.72}    & \multicolumn{1}{c|}{151.68}     &   30.08  \\ \cline{2-8} 
\multicolumn{1}{|c|}{}                     & total           & \multicolumn{1}{c|}{96.35}    & \multicolumn{1}{c|}{100.22}    &   132.95   & \multicolumn{1}{c|}{94.99}    & \multicolumn{1}{c|}{156.35}     &   31.66\\ \hline
\end{tabular}
\label{crosssenction}
\end{table}

Then three main sources of theoretical uncertainty are emphatically analyzed, the mass of charm quark, the renormalization scale, and the long-distance element matrix. First is the mass of the constituent charm quark. In Table \ref{mc}, the mass of the constituent charm quark is set to be $m_{c}=1.8\pm0.3$ GeV to analyze the theoretical uncertainty of cross section at $\sqrt{s}=500$ GeV, and $m_{c}$ takes five options, $m_c=1.5$, 1.65, 1.8, 1.95 and 2.1 GeV. Note that when analyzing the theoretical uncertainty arising from one source, the other parameters from different sources remain at their central values. Table \ref{mc} shows that with the increase of $m_{c}$, the cross section decreases significantly for the suppression of the phase space. 
Secondly, by typically changing the renormalization energy scale $\mu$ to be $\mu_0/2$, $\mu_0$, and $2\mu_0$ with $\mu_0 = \sqrt{4m_{c}^2 + p^2_T}$, the dependence of the cross section on the energy scale can be seen from Table \ref{scale}. Due to the asymptotic freedom characteristic of QCD, as the energy scale increases (decrease), the strong running coupling constant $\alpha_s$ decreases logarithmically, resulting in a decrease (increase) in the cross section. 
Finally, three hadronization schemes are applied to compare the nonperturbative effect of the long-distance matrix element ${\mathcal O}_{\langle cc\rangle[n] }^{ T_{cc}}$, which is presented in Table \ref{TLDME}. As can be seen from Table \ref{TLDME}, the contributions for the photoproduction of $T_{cc}[^3S_1]_{\bar{3}}$ in HOP hadronization scheme are much greater than those in DAS and HDAS hadronization scheme, being 2.28 times and 36.63 times respectively, because the long-distance element matrix obtained in HOP hadronization scheme are much larger, while the matrix elements obtained by HDAS are the smallest. This is also the reason why we select the matrix element obtained by HDAS as the center value for numerical calculation in Tables \ref{crosssenction}-\ref{scale}. This phenomenon shows a very large non-perturbation effect.

\begin{table}[htb]
\caption{Theoretical uncertainty of the cross sections (in unit: fb) by varying $m_c$=1.5, 1.65, 1.8, 1.95 and 2.1 GeV for the photoproduction of $T_{cc}[^3S_1]_{\bar{3}}$ at $\sqrt{s}=500$ GeV using WWA and LBS spectrum.}
\begin{tabular}{|cc|c|c|c|c|c|}
\hline
\multicolumn{2}{|c|}{$m_c$ (GeV)}                            & 1.5   & 1.65  & 1.8  & 1.95 & 2.1  \\ \hline
\multicolumn{1}{|c|}{\multirow{3}{*}{WWA}} & $\gamma+\gamma$ & 21.93 & 12.41 & 7.37 & 4.55 & 2.91 \\ \cline{2-7} 
\multicolumn{1}{|c|}{}                     & $\gamma$+g      & 1.54  & 0.83  & 0.47 & 0.28 & 0.17 \\ \cline{2-7} 
\multicolumn{1}{|c|}{}                     & total           & 23.47 & 13.24 & 7.84 & 4.83 & 3.08 \\ \hline
\multicolumn{1}{|c|}{\multirow{3}{*}{LBS}} & $\gamma+\gamma$ & 48.60 & 32.20 & 22.11 & 15.64 & 11.36 \\ \cline{2-7} 
\multicolumn{1}{|c|}{}                     & $\gamma$+g      & 227.79 & 130.27 & 78.11 & 48.72 & 31.30\\ \cline{2-7} 
\multicolumn{1}{|c|}{}                     & total           & 276.39 & 162.47 & 100.22 & 64.36 & 42.66 \\ \hline
\end{tabular}
\label{mc}
\end{table}

\begin{table}[htb]
\caption{Theoretical uncertainty of the cross sections (in unit: fb) by varying $\mu$=$\mu_0/2$, $\mu_0$ and $2\mu_0$ with $\mu_0=\sqrt{4m_{c}^{2}+p_{T}^{2}}$ for the photoproduction of $T_{cc}[^3S_1]_{\bar{3}}$ at $\sqrt{s}=500$ GeV using WWA and LBS spectrum.}
\begin{tabular}{|c|ccc|ccc|}
\hline
\multirow{2}{*}{Scale} & \multicolumn{3}{c|}{WWA}                                                 & \multicolumn{3}{c|}{LBS}                                                 \\ \cline{2-7} 
                  & \multicolumn{1}{c|}{$\mu_0/2$} & \multicolumn{1}{c|}{$\mu_0$} & $2\mu_0$ & \multicolumn{1}{c|}{$\mu_0/2$} & \multicolumn{1}{c|}{$\mu_0$} & $2\mu_0$ \\ \hline
$\gamma +\gamma$  & \multicolumn{1}{c|}{12.53}      & \multicolumn{1}{c|}{7.37}    & 5.02 & \multicolumn{1}{c|}{36.58}      & \multicolumn{1}{c|}{22.11}    & 15.31         \\ \hline
$\gamma$ + g      & \multicolumn{1}{c|}{0.58}      & \multicolumn{1}{c|}{0.47}    & 0.40        & \multicolumn{1}{c|}{87.16}      & \multicolumn{1}{c|}{78.11}    & 70.37   \\ \hline
total      & \multicolumn{1}{c|}{13.11}      & \multicolumn{1}{c|}{7.84}    & 5.42        & \multicolumn{1}{c|}{123.73}      & \multicolumn{1}{c|}{100.22}    & 85.68   \\ \hline
\end{tabular}
\label{scale}
\end{table}

\begin{table}[htb]
\caption{Theoretical uncertainty of the cross sections (in unit: fb) by appling three hadronization schemes, HOP, HDAS, and DAS, to abtain the long-distance matrix element ${\mathcal O}_{\langle cc\rangle[n] }^{ T_{cc}}$ for the photoproduction of $T_{cc}[^3S_1]_{\bar{3}}$ at $\sqrt{s}=500$ GeV using WWA and LBS spectrum.}
\begin{tabular}{|cc|c|c|c|}
\hline
\multicolumn{2}{|c|}{${\mathcal O}_{\langle cc\rangle[n] }^{ T_{cc}}$}                                           & HOP & HDAS & DAS \\ \hline
\multicolumn{1}{|c|}{\multirow{3}{*}{WWA}} & $\gamma+\gamma$ & 280.17 & 7.37 & 122.77 \\ \cline{2-5} 
\multicolumn{1}{|c|}{}                              & $\gamma$+g      & 17.89  & 0.47 & 7.84 \\ \cline{2-5} 
\multicolumn{1}{|c|}{}                              & total           & 298.06 & 7.84  & 130.61  \\ \hline
\multicolumn{1}{|c|}{\multirow{3}{*}{LBS}} & $\gamma+\gamma$ & 840.93 & 22.11 & 368.50 \\ \cline{2-5} 
\multicolumn{1}{|c|}{}                              & $\gamma$+g      & 2970.92 & 78.11 & 1301.86 \\ \cline{2-5} 
\multicolumn{1}{|c|}{}                              & total           & 3811.86 & 100.22 & 1670.36 \\ \hline
\end{tabular}
\label{TLDME}
\end{table}

\begin{figure}
\centering
\hspace{-0.50in}
    \includegraphics[scale=0.32]{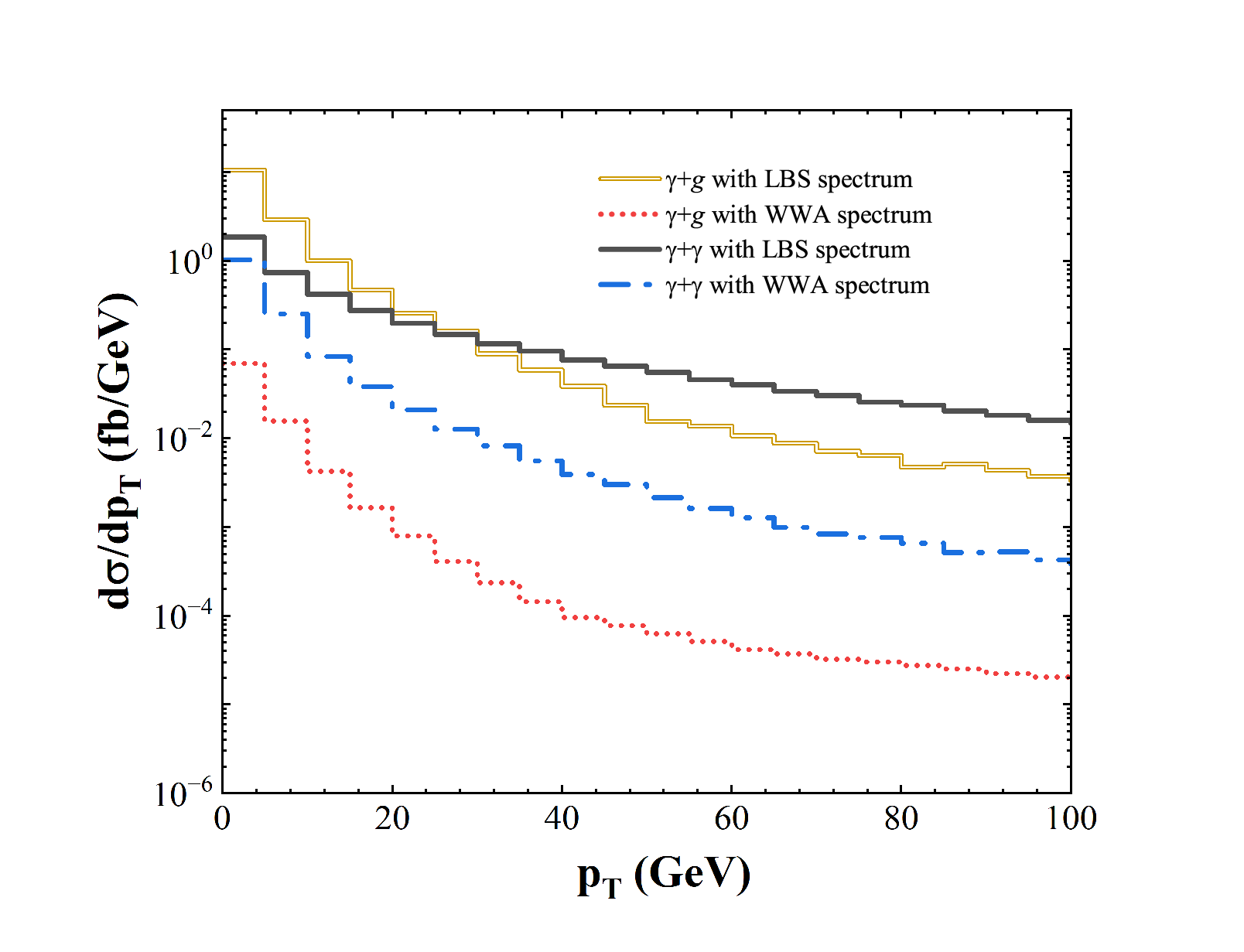}
  \hspace{-0.50in}
    \includegraphics[scale=0.32]{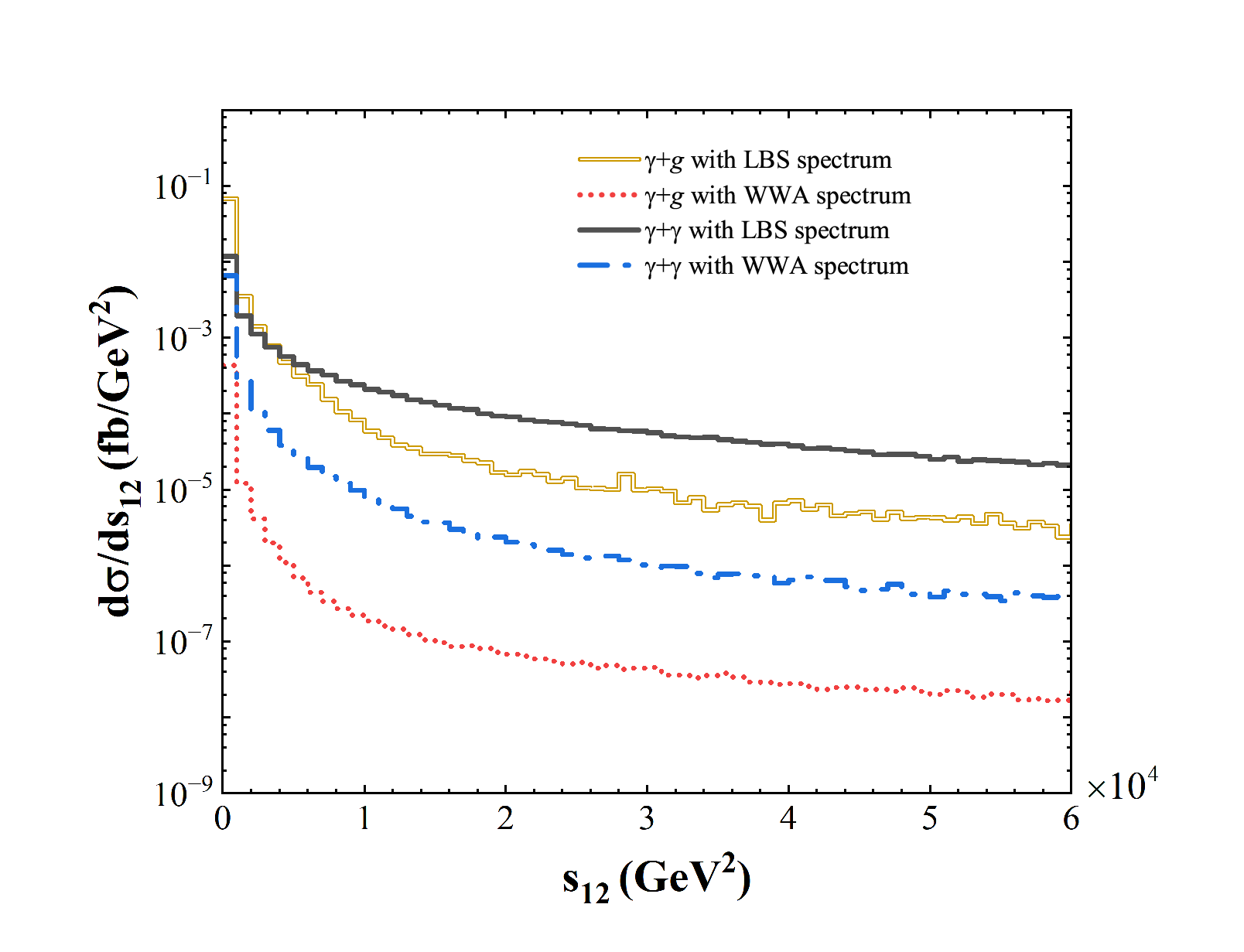}\\
\hspace{-0.50in}
    \includegraphics[scale=0.32]{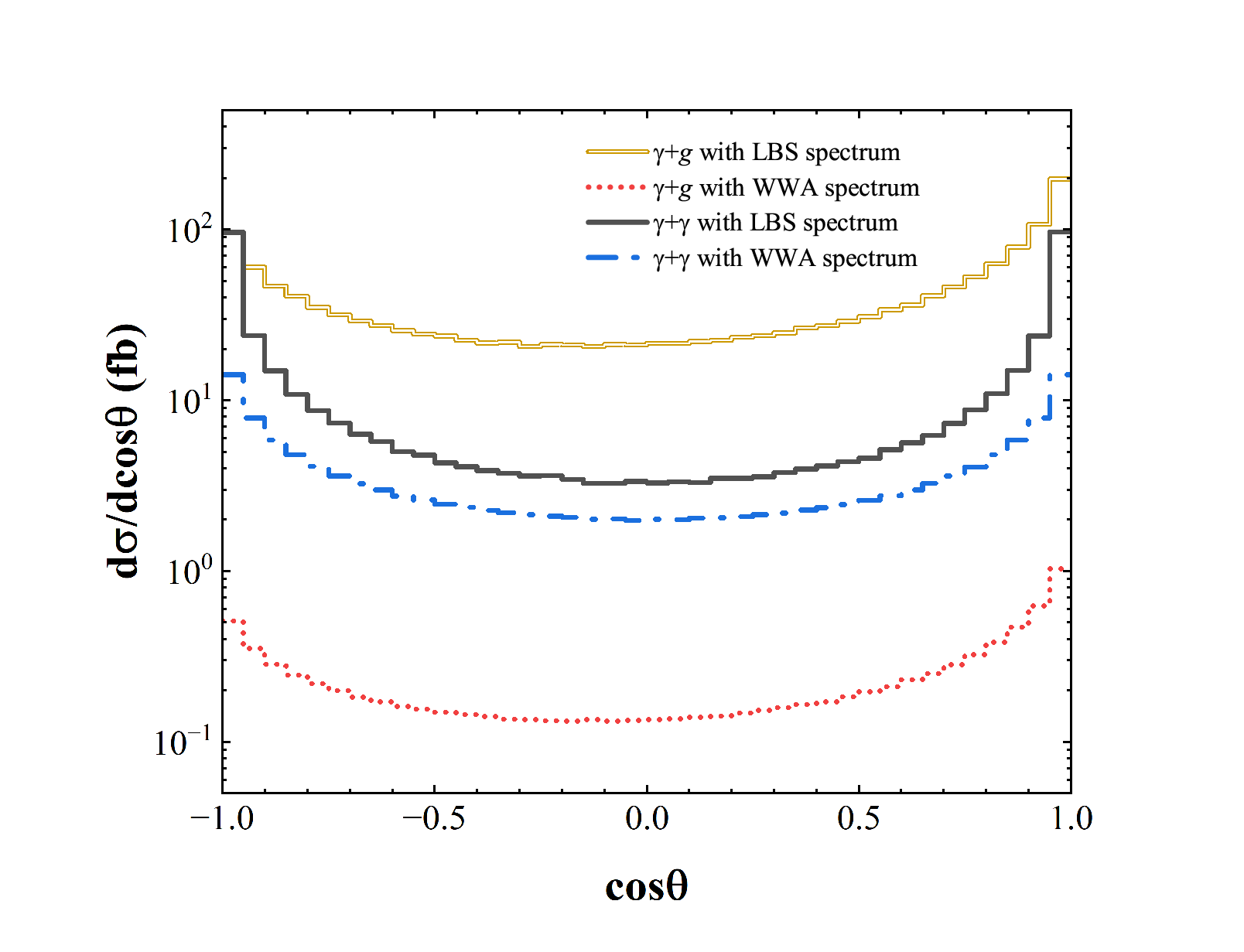}
\hspace{-0.50in}
    \includegraphics[scale=0.32]{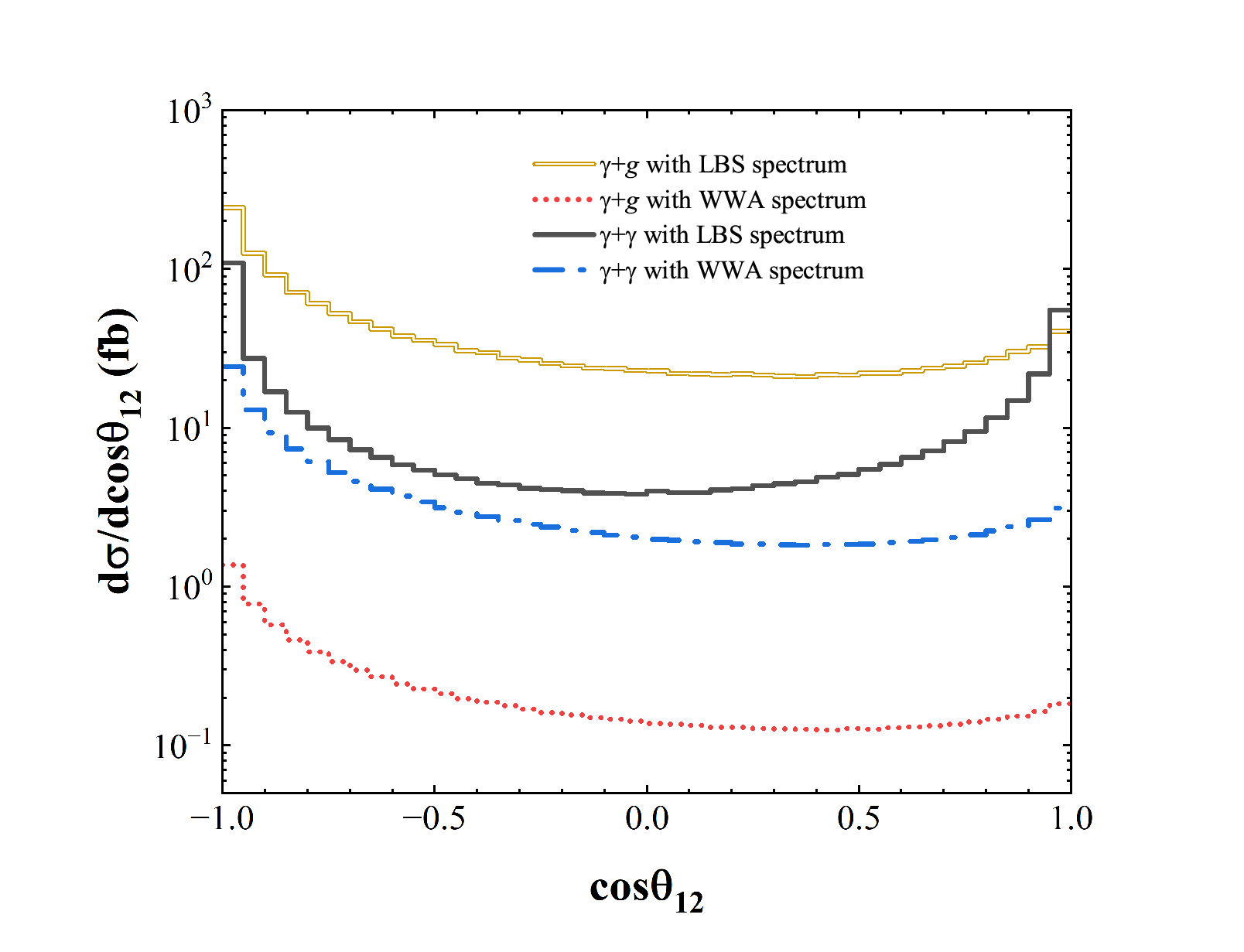}
\caption{Differential distributions for the photoproduction of $T_{cc}$ at $\sqrt{s}=500$ GeV.} 
\label{distributions}
\end{figure}

In order to show the kinematic properties and provide some guidance for future experimental detection, the differential distributions for the photoproduction of doubly charmed tetraquark $T_{cc}$ at $\sqrt{s}=500$ GeV are presented in Fig. \ref{distributions}, including the transverse momentum $p_T$ distribution, the invariant mass $s_{12}$, the angular cos$\theta$ and cos$\theta_{12}$, where the definition of squared invariant mass $s_{12} = (q_1 + q_2)^2$, $\theta$ ($\theta_{12}$) is the angle between the momenta $\vec{q_1}$ of $T_{cc}$ and the initial photon $\vec{p_1}$ (outgoing anticharmed quark $\vec{q_2}$). From Fig. \ref{distributions}, the contributions from LBS spectrum are much larger than that of using WWA. And in WWA spectrum, the contribution of photon-photon fusion (blue line) is larger than that of photon-gluon fusion (red line) in the whole region.
From the transverse momentum $p_T$ distribution and invariant mass $s_{12}$ distribution, the differential corss section shows a major downward trend for photoproduction of $T_{cc}$. And in LBS spectrum, the results from the resolved photon channel become dominant only in the small transverse momentum $p_t$ region, $p_t\le30$ GeV. From the angular distributions cos$\theta$ and cos$\theta_{12}$, it is shown that the curves present a valley shape. And the greatest contribution can be obtained when $\theta=0~(\pi)$, that is to say $T_{cc}$ is most likely to move back-to-back with the anticharm quark $\bar{c}$ in the beam direction.

Combined with the experimental parameters, different transverse momentum cuts, $p_T\geq1,~5, ~10$ GeV, are adopted to evaluate the cross sections for the photoproduction of $T_{cc}$ at ILC and CLIC. The results are listed in Table \ref{ptcut}. From Table \ref{ptcut}, we see that the cross sections for the photoproduction of $T_{cc}$ through LBS (WWA) with $p_T\geq1,~5, ~10$ GeV cut are 85.2\% (84.2\%), 38.4\% (29.8\%), 20.3\% (12.7\%)of that without the transverse momentum cut correspondingly. Therefore, the contributions in small $p_T$ region are crucial for experimental detection.

\begin{table}[htb]
\caption{Cross sections (in unit: fb) by added different transverse momentum cuts, $p_T\geq1,~5, ~10$ GeV, for the photoproduction of $T_{cc}[^3S_1]_{\bar{3}}$ at $\sqrt{s}=500$ GeV using WWA and LBS spectrum.}
\begin{tabular}{|cc|c|c|c|}
\hline
\multicolumn{2}{|c|}{$p_T$ cuts}                                           & $p_T\geq1$ GeV & $p_T\geq~5$ GeV & $p_T\geq10$ GeV \\ \hline
\multicolumn{1}{|c|}{\multirow{3}{*}{WWA}} & $\gamma+\gamma$ & 6.21 & 2.22 & 0.95 \\ \cline{2-5} 
\multicolumn{1}{|c|}{}                              & $\gamma$+g      & 0.39  & 0.12 & 0.04 \\ \cline{2-5} 
\multicolumn{1}{|c|}{}                              & total           & 6.60 & 2.34  & 0.99  \\ \hline
\multicolumn{1}{|c|}{\multirow{3}{*}{LBS}} & $\gamma+\gamma$ & 19.45 & 12.78 & 9.08 \\ \cline{2-5} 
\multicolumn{1}{|c|}{}                              & $\gamma$+g      & 65.97 & 25.72 & 11.31 \\ \cline{2-5} 
\multicolumn{1}{|c|}{}                              & total           & 85.42 & 38.50 & 20.39 \\ \hline
\end{tabular}
\label{ptcut}
\end{table}

%%%%%%%%%%%%%%%%%%%%%%%%%%%%%%%%%%%%%%%%%%
\section{SUMMARY}
\label{summary}

Within the framework of NRQCD, the photoproduction of doubly charmed tetraquark $T_{cc}$ has been analyzed through the resolved photon channel,  $\gamma+g\to \langle cc\rangle [n]+\bar{c}+\bar{c} \to T_{cc}+\bar{c}+\bar{c} $ at leading order in ${\cal O}(\alpha_s^3)$, where both WWA and LBS photon spectrum are adopted for the initial photon at ILC and CLIC. Only the color and spin quantum number $[^3S_1]_{\bar{3}}$ of the intermediate $\langle cc\rangle$-diquark is taken into consideration to form the compact $T_{cc}$.

The corss sections for the photoproduction of photon-gluon fusion and photon-photon fusion are numerically calculated at ILC with $\sqrt{s} = 250,~500,~1000\ {\rm GeV}$ and CLIC with $\sqrt{s} = 380,~1500,~3000\ {\rm GeV}$. From Table \ref{crosssenction}, it can be seen that the contribution from LBS photon spectrum is dominant than WWA. In WWA photon spectrum, as the CM energy increases, the contribution of the resolved photon-gluon channel continues to rise, being 3.53\%, 6.38\%, 12.61\% of that of the photon-photon channel. The same trend can also be obtained at CLIC.
While in LBS photon spectrum, the contribution of the resolved photon-gluon channel increased significantly with the increase of CM energy at ILC, up to 88.77\%, 3.53, 14.60 times of that of the photon-photon channel. At CLIC, the contribution of the resolved photon-gluon channel can be 2.04, 32.48, and 19.04 times of that of the photon-photon channel at $\sqrt{s}=$380, 1500, and 3000 GeV.
Therefore, the study of the production of tetraquark states through resolved photon-gluon channel cannot be ignored at $e^+e^-$ collider.

With the designed luminosity $\mathcal L=2, 4, 8~ {\rm ab}^{-1}$ corresponding to ILC at $\sqrt{s}=$ 250, 500, and 1000 GeV and $\mathcal L=1, 2.5, 5~{\rm ab}^{-1}$ for CLIC at $\sqrt{s}=$380, 1500, and 3000 GeV, respectively, the total produced events of $T_{cc}$ via photoproduction can be approximately $9.37\times 10^3- 8.43\times 10^4$ ($1.92\times 10^5- 1.06\times 10^6$) each year at ILC with $\sqrt{s}=$ 250, 500, and 1000 GeV using WWA (LBS) photon spectrum. At CLIC with $\sqrt{s}=$380, 1500, and 3000 GeV, the total produced events of $T_{cc}$ via photoproduction can be $6.58\times 10^3- 6.57\times 10^4$ ($9.50\times 10^4- 3.91\times 10^5$) each year using WWA (LBS) photon spectrum. Based on such a base number and Considering the subsequent decay of the doubly charmed tetraquark, it is still possible for $T_{cc}$ to be experimentally discovered at linear electron-positron colliders through photoproduction.

Three main sources of theoretical uncertainty, the mass
of charm quark, the renormalization scale, and the long-distance element matrix, are emphatically analyzed, which are shown in Tables \ref{mc}-\ref{TLDME}. The constituent charm quark mass of $(cc)$-diquark is set to be $m_{c}=1.8\pm0.3$ GeV and the renormalization energy scale $\mu$ is changed to be $\mu_0/2$, $\mu_0$, and $2\mu_0$ with $\mu_0 = \sqrt{4m_{c}^2 + p^2_T}$ to analyze the theoretical uncertainty. Three different hadronization schemes, HOP, HDAS, and DAS for the $(cc)[^3S_1]_{\bar{3}}$ into $T_{cc}$ are also discussed. This phenomenon shows a very large non-perturbation effect. 

Finally, the transverse momentum, the invirant mass, and the angular distributions of $T_{cc}$ are presented in Fig. \ref{distributions}. As can be found in LBS spectrum, the results from the resolved photon channel become dominant only in the small transverse momentum $p_t$ region, $p_t\le30$ GeV. From the angular distributions cos$\theta$ and cos$\theta_{12}$, it is shown that the greatest contribution can be obtained when $\theta=0~(\pi)$, that is to say $T_{cc}$ is most likely to move back-to-back with the anticharm quark $\bar{c}$ in the beam direction.

%%%%%%%%%%%%%%%%%%%%%%%%%%%%%%%%%%%%%%%%%%%%%%%%%%%%%%%%%%%%%%%%%%%%%
\vspace{1.4cm} {\bf Acknowledgments}
Thanks for helpful disscusion with Xi-Jie Zhan and Jun Jiang. This work was partially supported by the Natural Science Foundation of Guangxi (no. 2024GXNSFBA010368, 2025GXNSFAA069775). This work was also supported by the Young Elite Scientists Sponsorship Program by GXAST (no. 2025YESSGX005).

%%%%%%%%%%%%%%%%%%%%%%%%%%%%%%%%%%%%%%%%%%%%%%%%%%%%%%%%%%%%%%%%%%%%

\end{document}